\documentclass[9pt]{IEEEtran}

\usepackage{color}
\usepackage{spconf}
\usepackage{times}

\usepackage{color}
\usepackage{theorem}
\usepackage{amsmath}
\usepackage{amssymb}
\usepackage{bm}
\usepackage{subfigure}
\usepackage{cite}
\usepackage{hyperref}

\usepackage[font=small,format=plain,up,up]{caption}
\usepackage{graphicx}
\usepackage{needspace}

\input{mysymbol.sty}

\title{Enhancing Geometric Deep Learning via Graph Filter Deconvolution}

\name{Jingkang Yang$^{a,b}$ and Santiago Segarra$^{c}$
	\thanks{SS received an IDSS MIT seed grant. 
	Emails: jk.yang1995@gmail.com, segarra@mit.edu}
\address{$^{a}$ School of Electronic Engineering and Computer Science, Queen Mary University of London, London, UK\\
	$^{b}$ International School, Beijing University of Posts and Telecommunications, Beijing, China\\
	$^{c}$ Institute for Data, Systems, and Society, Massachusetts Institute of Technology, Cambridge, MA, USA}}
\begin{document}
\maketitle
\begin{abstract}
	In this paper, we incorporate a graph filter deconvolution step into the classical geometric convolutional neural network pipeline. 
	More precisely, under the assumption that the graph domain plays a role in the generation of the observed graph signals, we pre-process every signal by passing it through a sparse deconvolution operation governed by a pre-specified filter bank. 
	This deconvolution operation is formulated as a group-sparse recovery problem, and convex relaxations that can be solved efficiently are put forth. 
	The deconvolved signals are then fed into the geometric convolutional neural network, yielding better classification performance than their unprocessed counterparts.
	Numerical experiments showcase the effectiveness of the deconvolution step on classification tasks on both synthetic and real-world settings.
\end{abstract}

\section{Introduction}\label{S:intro}
Graph signal processing (GSP) emerges in response to the need to better process and understand the ever-increasing volume of network data, often conceptualized as signals defined on graphs \cite{shuman2013emerging, Sandryhaila2013}. For example, graph-supported signals can model economic activity observed over a network of production flows between industrial sectors~\cite{marques_sampling_2016}, as well as brain activity signals supported on brain connectivity networks \cite{bullmore2009complex, huang_brain_2018, medaglia_brain_2017}. 
However, due to the complexity and irregularity of such networks, most of the standard signal processing notions -- such as convolution and downsampling -- are no longer directly applicable on graph settings. 
Fortunately, in the past years, numerous works have expanded several of the classical signal processing tools onto the realm of graphs, including sampling and reconstruction~\cite{marques_sampling_2016, chen_sampling_2015, segarra_reconstruction_2016, romero_reconstruction_2017}, stationarity and power spectral density estimation~\cite{marques2016stationaryTSP16, perraudinstationary2016, girault_stationarity}, filter design~\cite{segarra_filters_2017, isufi_filters_2017}, and blind deconvolution~\cite{segarra2016blind, ramirez2017graph, ye2018blind}.
Of special importance to this paper is the latter body of work on (blind) deconvolution, where the input to a network diffusion process (formalized as a graph filter) is recovered from the observation of the corresponding output, thus extending deconvolution of temporal and spatial signals to the less structured graph domain.

General learning tasks for network data, such as classification and regression, are also of practical importance. 
For example, one might want to classify brain activity patterns while taking into account the brain network on which they are defined.
This provides an opportunity for a synergistic relation between machine learning and GSP, something currently being investigated under the name of geometric (deep) learning (GDL).
The general goal of GDL is to generalize structured deep neural networks to non-Euclidean domains, such as graphs and manifolds \cite{masci2016geometric}. 

GDL has achieved state-of-the-art performance on several application areas including shape correspondence tasks \cite{yi2017syncspeccnn, litany2017deep} and recommender systems \cite{monti2017geometric, huang_recommender_2017}. 
Different from classic convolutional neural networks (CNNs) that operate on well-structured data such as audio and images, GDL acts on irregular domains -- represented by graphs -- where the fundamental operations of convolution and pooling (downsampling) are not straightforward \cite{bronstein2017geometric}. 
Thus, most of the existing efforts are geared towards developing alternatives for these operations on general graphs.
To extend the classic pooling operation, most works rely on existing clustering techniques to coarsen the graph structures \cite{von2007tutorial,dhillon2007weighted} while some recent approaches seek to avoid this operation in general graphs~\cite{gama_cnn_2018}. 
For the generalization of the convolutional layer, various classes of graph filters have been used, starting from full non-parametric definitions in the frequency domain \cite{bruna2013spectral} to a myriad of parametric definitions that simultaneously reduce overfitting and speed-up the training process \cite{defferrard2016convolutional, atwood2016diffusion, kipf2016semi}.

There is a clear potential for cross-pollination between GSP and GDL since the former is designed to process data defined on graphs whereas the latter seeks to better learn from this data. However, most of the attention so far has been dedicated to the use of GSP tools to design graph filters that better mimic the convolutional layer in traditional CNNs. We contend that there is room for further collaboration between these fields, where concepts like deconvolution of graph signals can be used for classification tasks.

{\bf Contribution and paper organization.}
In this paper, we incorporate the notion of graph filter deconvolution into the pipeline of GDL. 
More specifically, every signal to be classified is first passed through a deconvolution block using a pre-specified filter bank with sparse regularizers that jointly promote sparsity in the number of filters used in the deconvolution and the number of non-zero elements in the deconvolved (seeding) signals.
The classification is then performed on the seeding signals, leading to better performance in practice. 

The rest of the paper is organized as follows. In Section~\ref{S:graphs_background} we briefly review basic notions of GSP and GDL.
In Section~\ref{S:graphs_method} we present the sparse deconvolution problem and how it can be incorporated into the traditional GDL pipeline.
Section~\ref{S:num_exp} illustrates the performance benefits of the proposed approach in the classification of both synthetic and real-world data, before concluding in Section~\ref{S:conclusions} with closing remarks and potential avenues for future research.

\section{Background}\label{S:graphs_background}
\subsection{Fundamentals of Graph Signal Processing}\label{Ss:gsp}

Consider the (possibly directed) graph $\mathcal{G} = (\ccalN,\ccalE)$ formed by the set $\ccalN$ of $N$ nodes and the set of edges $\mathcal{E}$, such that the pair $(i,j)$ belongs to $\ccalE$ if there exists a link from node $i$ to node $j$. 
Associated with a given $\ccalG$, a graph signal can be represented as a vector $\bbx = [x_1, \ldots, x_N]^T \in \reals^{N}$, where the $i$th component, $x_i$, represents the signal value at node $i$. 
The network structure is captured by the graph-shift operator $\bbS$ \cite{Sandryhaila2013}, a sparse matrix such that $[\bbS]_{ij} \neq 0$ for $(i,j) \in \mathcal{E}$ or $i = j$. 
The adjacency matrix and the graph Laplacian are usual choices for the shift operator. 
Assuming that $\bbS$ is diagonalizable, the shift can be decomposed as $\bbS = \bbV \diag(\bblambda) \bbV^{-1}$, where $\bblambda=[\lambda_1,\ldots,\lambda_N]^T$ collects the eigenvalues. 
Linear graph filters are defined as graph-signal operators of the form $\bbH = \sum_{l = 0}^{L-1} h_l \bbS^l,$ i.e., polynomials in $\bbS$ \cite{Sandryhaila2013}. 
The filtering operation is thus given by $\bby = \bbH \bbx$, where $\bby$ is the filtered signal, $\bbx$ the input, $\bbh=[h_0, \ldots, h_{L-1}]^T$ the filter coefficients, and $L-1$ the filter degree. Notice that $\bby$ is effectively a linear combination of successive shifted versions of the input $\bbx$. Indeed, graph filters are the natural extension of convolutions to the graph domain.

The filter coefficients $\bbh$ determine the range of influence of the filter in the graph domain. For example, if for a given filter only the first three elements of $\bbh$ are significant, then the output of the filter at node $i$ will only be determined by the input values at node $i$ and its two-hop neighborhood. 
In this paper we consider four types of filters with varying range: i) uniform-range $\bbh_{\text{UR}}$ where $\bbh_{\text{UR}} = \mathbf{1}$, ii) short-range $\bbh_{\text{SR}}$ where $[\bbh_{\text{SR}}]_i = (1 + i/L)^{-4}$, iii) medium-range $\bbh_{\text{MR}}$ where $[\bbh_{\text{MR}}]_i = \exp({-(i/L - 0.5)^2})$ , and iv) long-range $\bbh_{\text{LR}}$ where $[\bbh_{\text{LR}}]_i = [\bbh_{\text{SR}}]_{L-i+1}$. 
As expected, the values of $[\bbh_{\text{SR}}]_i$ decrease with $i$ whereas the opposite is true for the values of $[\bbh_{\text{LR}}]_i$. Moreover, for the medium-range filter, the values of $[\bbh_{\text{MR}}]_i$ are given by a Gaussian kernel centered at $L/2$.

As in classical signal processing, graph filters and signals may be represented in the frequency (or Fourier) domain. 
Defining the graph Fourier operator as $\bbU = \bbV^{-1}$, the graph Fourier transform (GFT) of the signal $\bbx$ is $\tilde{\bbx} = \bbU \bbx$. 
For graph filters, the definition of the GFT that maps the filter coefficients $\bbh$ to the frequency response of the filter, $\tbh$, is given by $\tilde{\bbh} = \bbPsi \bbh$, where $\bbPsi$ is an $N \times L$ Vandermonde matrix whose elements are $[\bbPsi]_{i,j} = \lambda_i^{j-1}$ \cite{segarra_filters_2017}. 
With $\circ$ denoting the Hadamard (elementwise) product, the definitions of the GFT for signals and filters allow us to rewrite the filtering operation in the spectral domain as $\tilde{\bby} = \tilde{\bbh} \circ \tilde{\bbx}$, mimicking the classical convolution theorem.

\subsection{Basics of Geometric Deep Learning}\label{Ss:GDL}

The main goal of GDL is to generalize structured deep neural networks to non-Euclidean domains, such as graphs and manifolds. 
Specifically, generalizing CNNs for data defined in irregular domains is one of the main challenges of GDL.
In this section, we review some of the existing approaches to overcome this challenge. We separate the presentation into the three main components of CNNs: convolution, non-linearity, and pooling.

{\bf Convolutional layer.} Graph filters naturally extend the convolution operation to general graphs (cf. Section~\ref{Ss:gsp}).
Hence, for a given graph-shift operator, to represent a convolutional layer with $R$ input channels (each input represents a graph signal) and $M$ output channels, we use the following expression
\begin{equation}\label{E:SCNN}
{\bf g}^{(m)} = \bbV \sum_{r=1}^{R} \diag(\tbh^{(m,r)}) \bbV^{-1} {\bf f}^{(r)}, \quad \text{for} \,\, m=1, \ldots, M, 
\end{equation}
where $( {\bf f}^{(1)}, \dots, {\bf f}^{(R)})$ represent the $R$ input channels, $(\bbg^{(1)}, \ldots, \bbg^{(M)})$ represent the $M$ output channels and $\tbh^{(m,r)} \in \reals^{N \times 1}$ is the frequency response of one of the $M \times R$ filters that encode the convolutional layer.
These frequency responses are the ones learned during the training process.
We denominate this method `non-parametric Geometric CNN (non-parametric GNN)', since every element of every frequency response $\tbh^{(m,r)}$ is decoupled and can be tuned freely.
To prevent overfitting and to promote locality of the learned filters, one can enforce the frequency responses $\tbh^{(m,r)}$ to belong to some parametric subspace~\cite{litany2017deep}, as the one described by cubic splines.
Specifically, we can define the responses $\tbh^{(m,r)} = \bbB \bbalpha^{(m,r)}$, where $\bf B$ is an $N \times Q$ fixed interpolation kernel encoding cubic splines, and  $\bbalpha^{(m,r)} \in \reals^{Q \times 1}$ is a vector of learnable parameters\cite{bruna2013spectral}.
We name this method as `spline GNN'.
Additional parametric families of graph filters have been defined in the literature~\cite{defferrard2016convolutional, atwood2016diffusion, kipf2016semi}.

{\bf Non-linearity.} 
Activation functions $\xi$ are non-linear functions applied elementwise to the outputs of convolutional layers before the action of the pooling layer. 
Given the elementwise application of $\xi$, activation functions can be implemented in a straightforward manner for GDL. In this paper we use the well-known rectified linear unit (ReLU) activation function $\xi(x)= \max(0,x)$.

{\bf Pooling layer.}
The pooling layer reduces the dimensionality of the signal being treated. For the treatment of images by classical CNNs, this operation is simple where groups of neighboring pixels are merged into single pixels, thus obtaining a smaller image as the output of the layer. However, for the pooling layer in irregular graphs one has to group nodes into blocks via clustering operations and then define a graph topology between the coarser nodes.
Although multiple methods co-exist in the literature, in this paper we use the fast pooling method proposed in~\cite{defferrard2016convolutional}, which makes use of the Graclus coarsening algorithm~\cite{dhillon2007weighted} to build a binary tree structure for the pooling layer.


\begin{figure*}
	\centering
	\includegraphics[width=6cm]{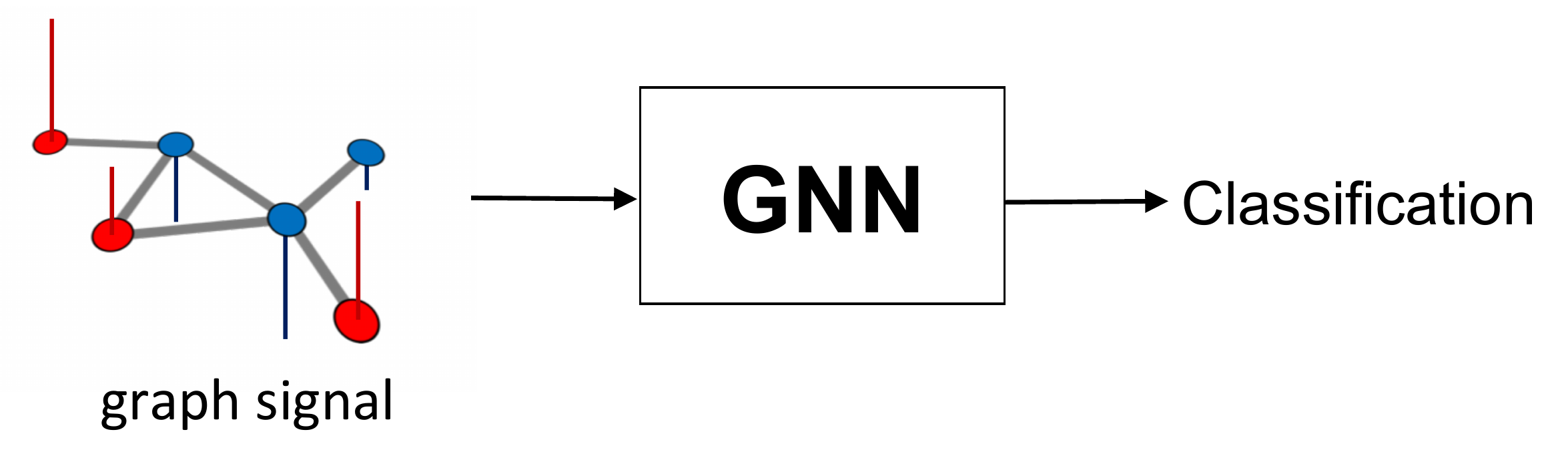}
	\hspace{0.5cm}
	\includegraphics[width=10.5cm]{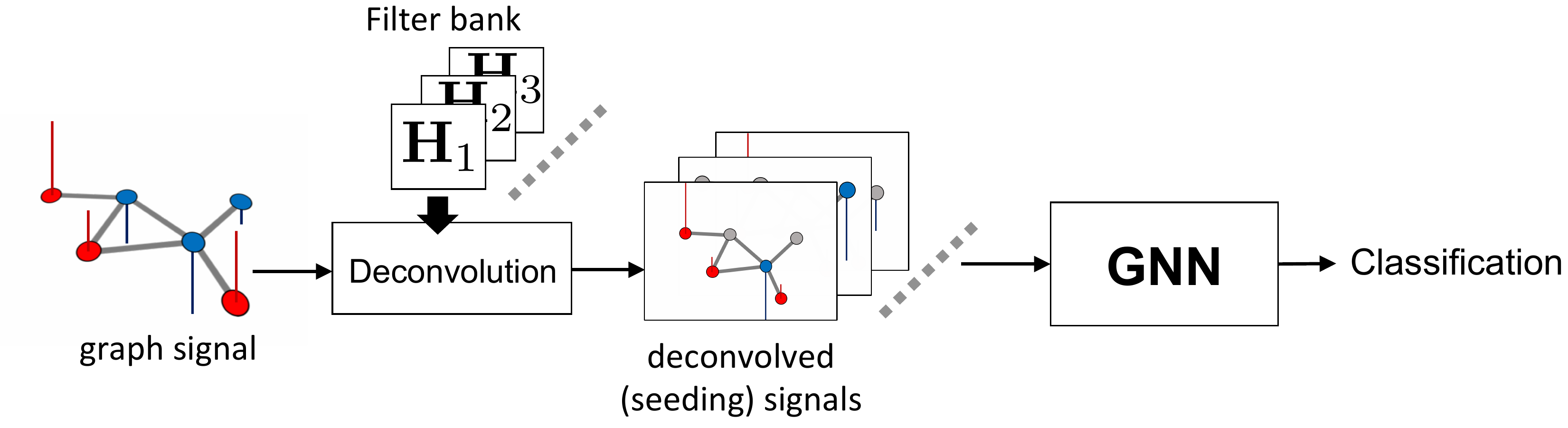}\\
	\hspace{0.5cm} (a) \hspace{8cm} (b) \hspace{2.5cm}
	\caption{Sketch of the classical and proposed GDL pipelines. (a) In the classical approach, a given graph signal is passed through a GNN to obtain a classification output. (b) We propose to first pass the given signal through a deconvolution block using a pre-specified filter bank. The GNN is then applied to the seeding (deconvolved) signals.}
	\vspace{-0.2cm}
	\label{fig:classic_and_proposed}
\end{figure*}

\section{Graph Filter Deconvolution for GDL}\label{S:graphs_method}

The archetypal formulation of a GDL classifier consists of a GNN whose weights are learned during a training phase via backpropagation, and then can be readily used as in Fig.~\ref{fig:classic_and_proposed}(a). More precisely, whenever a new graph signal has to classified, it is processed by the trained GNN and a classification output is obtained. This simple pipeline processes the graph signals as given while relying on the graph convolutions to generate useful, graph-dependent features for classification.
However, this pipeline ignores potential generative methods of the graph signal observed in the first place. 
In particular, it is often the case that graph signals are obtained from some diffusion process in the graph, for example, the contagion of an epidemic or the spread of news. Moreover, if this diffusion process is common across the classes that we are trying to differentiate, it then obscures the difference between classes, complicating the task of the GNN.
In this context, we propose the alternative pipeline in Fig.~\ref{fig:classic_and_proposed}(b). Instead of directly using graph signals as inputs of the GNN, we first perform a deconvolution step with a pre-specified filter bank to obtain for each signal a set of sparse (deconvolved) seeding signals. 
These seeding signals are then fed into the GNN for either training or classification. 
In the remainder of this section we elaborate on the deconvolution step.

\begin{figure*}
	\centering
	\includegraphics[width=6cm]{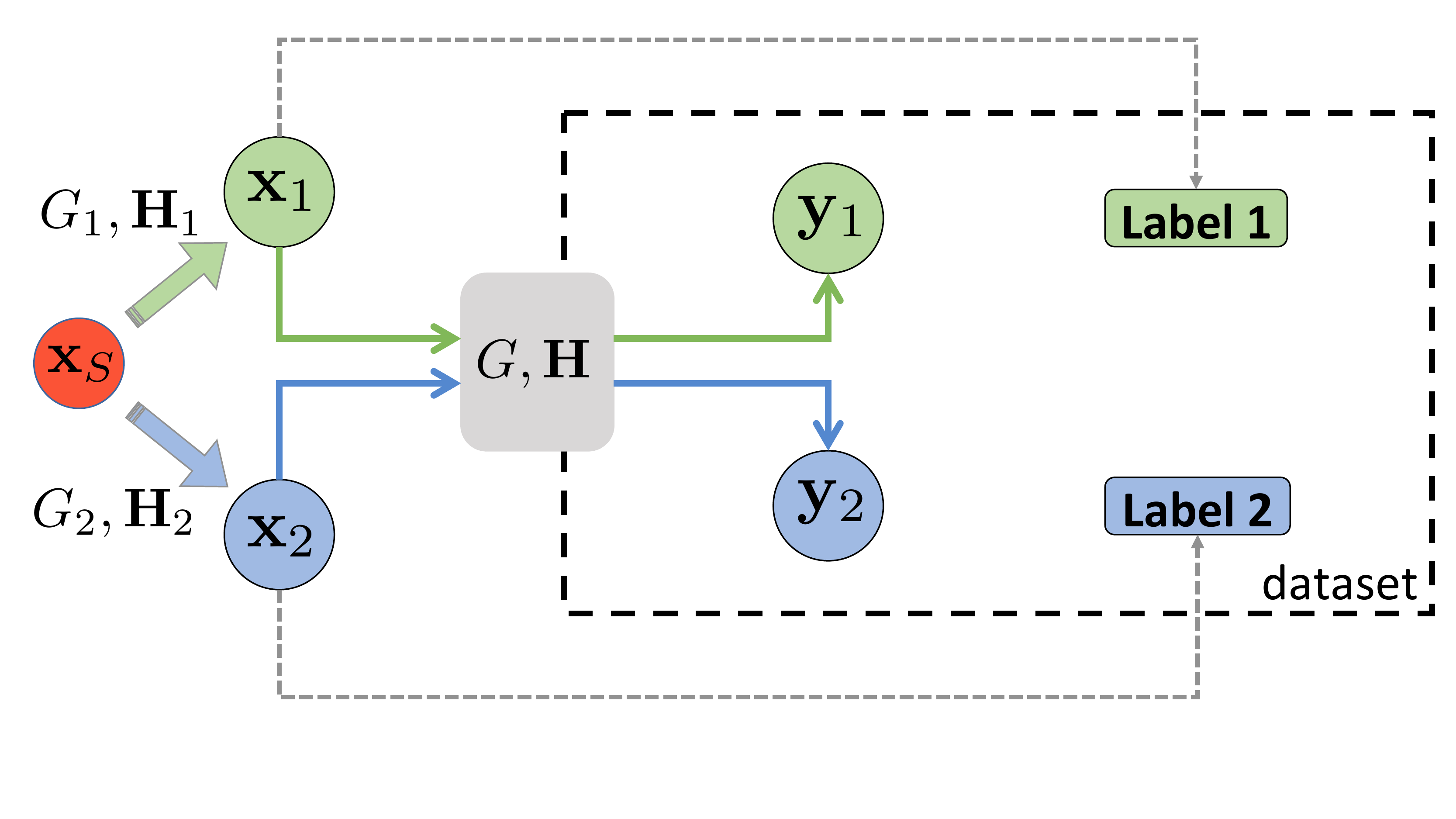} \hspace{0.3cm}
	\includegraphics[width=5cm]{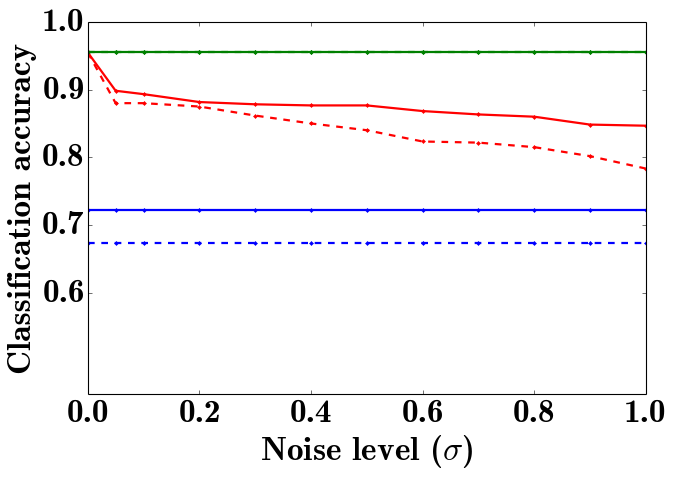}  \hspace{0.3cm}
	\includegraphics[width=5cm]{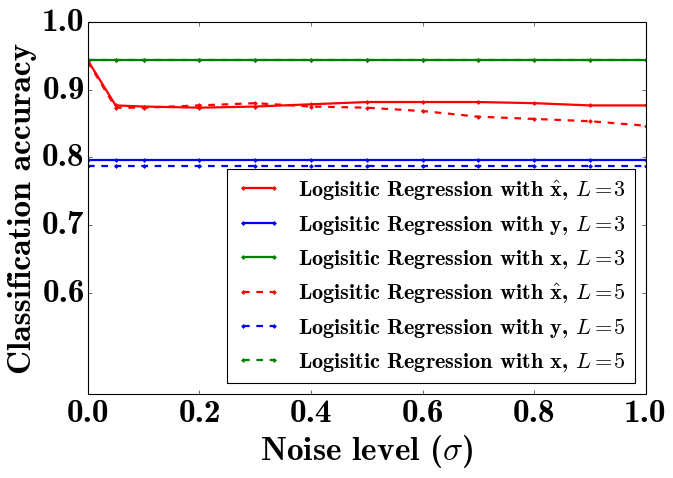}\\
	\hspace{1cm} (a) \hspace{5.5cm} (b) \hspace{5cm} (c)  
	\caption{(a) Schematic view of the generation process for the synthetic data. (b) Classification accuracy as a function of the noise level $\sigma$ in the deconvolution filter for three different classifiers based on $\bbx$ (green), $\bby$ (blue), and $\hbx$ (red). The solid and dashed lines respectively correspond to $L=3$ and $L=5$ for the diffusing filter. The size of the graphs is $N=30$. (c) Counterpart of (b) for $N=50$.}
	\vspace{-0.2cm}
	\label{fig:sketch}
\end{figure*}

Consider the case where, indeed, the observed graph signals $\bby$ (such as the current state of an epidemic) can be modeled as the output of a diffusion process $\bbH$ excited by an originally sparse input $\bbx$ (such as a small group of `patients zero'). Formally, we have that $\bby = \bbH \bbx$, where the seeding signal $\bbx$ is sparse with at most $S \ll N$ non-zero entries. 
Whenever $\bbH$ is known, one can recover $\bbx$ from $\bby$ by solving the sparse recovery problem
\begin{equation}\label{E:decon_single_filter}
\min_{\bbx} \| \bbx \|_1 \qquad \text{subject to} \quad \bby = \bbH \bbx, 
\end{equation}
where the $\ell_1$ norm in the objective function is used as the usual convex surrogate of the sparsity promoting $\ell_0$ (pseudo-)norm. Naturally, whenever the observed signal $\bby$ is noisy or the knowledge on $\bbH$ is not perfect, one may replace the constraint in \eqref{E:decon_single_filter} by its robust counterpart $\| \bby - \bbH \bbx \|_2 \leq \epsilon$, where $\epsilon$ is tuned according to some prior knowledge on the noise level.

A more challenging setting is one where neither $\bbH$ nor a noisy version of it are known. 
A possible approach in this situation is to implement the existing results on \emph{blind} graph filter identification~\cite{segarra2016blind, ramirez2017graph, ye2018blind} to jointly recover $\bbH$ and $\bbx$ from $\bby$.
However, solving such blind identification problems can be computationally demanding for graphs of size in the order of a thousand nodes.
To overcome this computational difficulty, we instead pre-specify a bank of filters with the intention that one of them (or a sparse combination of these) would be close to the true underlying diffusion filter.
Formally, consider a filter bank $\bbP = [{\bf H}_1, {\bf H}_2, \dots, {\bf H}_K] \in \mathbb{R}^{N \times NK}$ that collects $K$ possible diffusion processes for the graph signals at hand. In practice, we build $\bbP$ by combining filters of different range (cf.~Section~\ref{Ss:gsp}). 
Associating a seeding signal $\bbx_i$ with each of these filters, we define the matrix $\bbX = [\bbx_1, \ldots, \bbx_K] \in \reals^{N \times K}$.
With this notation in place, the proposed deconvolution step boils down to solving the following optimization problem
\begin{equation}\label{E:decon_multi_filter}
\min_{\bbX} \| \bbX \|_{2,1} + \alpha \|\text{vec}(\bbX)\|_{1}  \qquad \text{subject to} \quad \bby = \bbP \text{vec}(\bbX).
\end{equation}
The objective function in \eqref{E:decon_multi_filter} consists of two different regularizers, with $\alpha>0$ trading off their relative importance. The $\ell_{2,1}$ norm promotes column-sparsity in $\bbX$, i.e., that only a few of the columns of $\bbX$ are different from zero. This implies that we seek to explain the observed signal $\bby$ by using only a small number of the filters included in the bank $\bbP$. Furthermore, the $\ell_1$ norm, just like in \eqref{E:decon_single_filter} promotes that the few seeding signals that explain $\bby$ are themselves sparse. 
Notice that both problems \eqref{E:decon_single_filter} and \eqref{E:decon_multi_filter} are convex and, thus, can be solved efficiently in practice. This enables the implementation of the proposed GDL pipeline in Fig.~\ref{fig:classic_and_proposed}(b), as we illustrate in the next section.

\section{Numerical experiments}\label{S:num_exp}

The proposed method is evaluated on both synthetic data and the well-known MNIST handwritten digit dataset\cite{lecun1998gradient}. 
The results show that our deconvolution step improves both the accuracy and the learning rate of classification tasks.

\subsection{Synthetic data: Diffused graph signals}

We evaluate the benefits of deconvolution via a synthetic binary classification task with known ground truth. In order to generate the data, we follow the ensuing procedure.
We start with a sparse signal $\bbx_S$ of size $N \in \{30,50\}$ (we vary the size for different experiments) where only $10\%$ of the values in $\bbx_S$ are different from zero, and drawn from a standard normal distribution.  
Signal $\bbx_S$ is then separately processed by two filters $\bbH_1$ and $\bbH_2$ defined on two different graphs $G_1$ and $G_2$, drawn from Erd\H{o}s-R\'enyi (ER) random models with edge probabilities $p_1=0.1$ and $p_2 = 0.3$, respectively. 
Filter $\bbH_1$ is of uniform-range whereas filter $\bbH_2$ is of short-range (cf.~Section~\ref{Ss:gsp}). 
We denote the outputs of these filters as $\bbx_1$ and $\bbx_2$, respectively belonging to classes $1$ and $2$. Our observations are $\bby_1$ and $\bby_2$, obtained by diffusing $\bbx_1$ and $\bbx_2$ on the same graph $G$ (ER, $p=0.7$) via the same graph filter $\bbH$ (short-range of degree $L \in \{3,5\}$). Our objective is to classify $\bby_1$ and $\bby_2$ into their correct classes; see Fig.~\ref{fig:sketch}(a) for details.

We consider $1,\!000$ such sparse signals $\bbx_S$, leading to $2,\!000$ observed signals $\bby$ (half of each class), and we use $1,\!500$ for training and the remaining $500$ for testing.
We compare the classification accuracy of three different logistic regression classifiers: one based on the observed diffused signals $\bby$, another based on the true seeding signals $\bbx$, and the last one based on the deconvolved signals $\hat{\bbx}$. To obtain $\hat{\bbx}$ from $\bby$ we solve \eqref{E:decon_single_filter}. 
We do not assume perfect knowledge of $\bbH$ but rather we have access to a noisy version $\tilde{\bbH} = \bbH + \epsilon \bbZ$ where $\epsilon\sim\mathcal{N}(0,\,\sigma^{2})$ and every entry of $\bbZ$ is drawn from a standard normal distribution. In Figs.~\ref{fig:sketch}(b)-(c) we present the accuracy of these three classifiers as a function of $\sigma$. As expected, the classifier based on the true signal $\bbx$ (depicted in green) achieves the highest accuracy. By contrast, the classifier based on $\bby$ (portrayed in blue) suffers since the diffusion filter $\bbH$ obscures the differences between both classes. Lastly, the accuracy of the classifier based on the deconvolved signals $\hat{\bbx}$ (in red) largely depends on $\sigma$. When $\sigma = 0$, the perfect knowledge of $\bbH$ leads to noiseless deconvolution so that $\hat{\bbx} = \bbx$. As $\sigma$ increases, the performance degrades but is still preferable to the case where no deconvolution is performed. This effect is observed for different values of the filter degree $L$ and graph size $N$.

\begin{figure*}
	\centering
	\includegraphics[height=3.6cm]{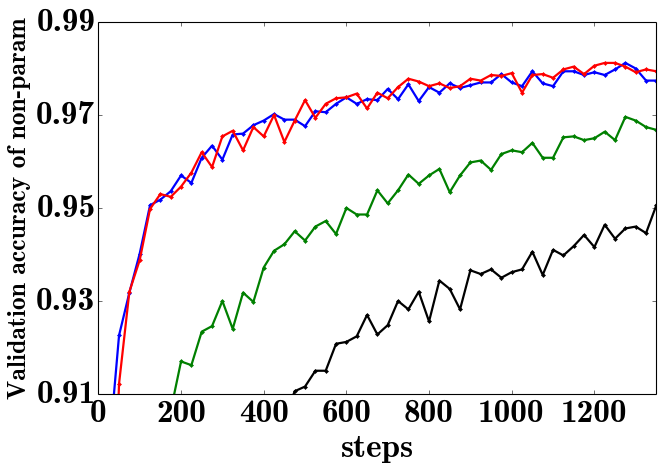} \hspace{0.3cm}
	\includegraphics[height=3.6cm]{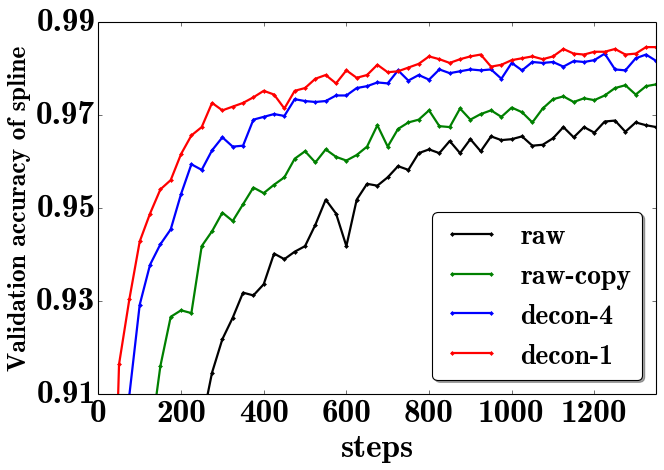} \hspace{0.3cm}
	\includegraphics[height=3.6cm]{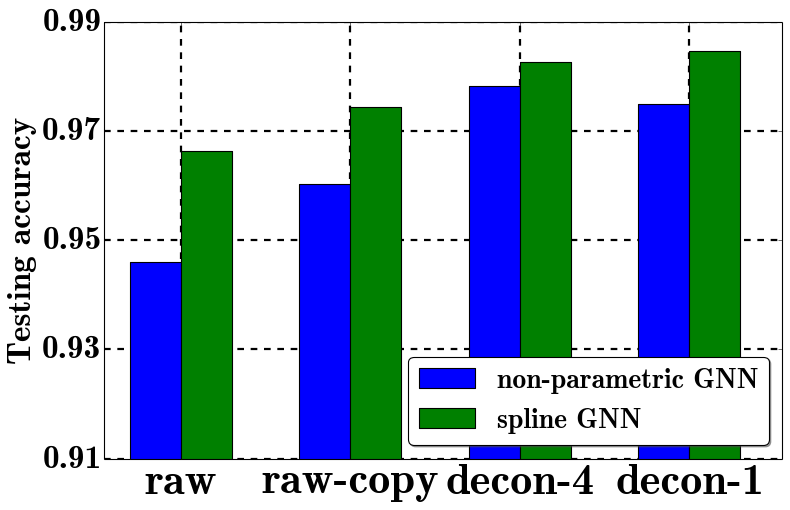}\\
	\hspace{1.8cm} (a) \hspace{5cm} (b) \hspace{5.3cm} (c)  \hspace{1.3cm}
	\caption{(a) Validation accuracy of the non-parametric GNN as a function of the training steps for the four classes of input signals considered. (b) Counterpart of (a) for the spline GNN. (c) Testing accuracy for both GNNs and the four classes of input signals considered.}
	\vspace{-0.2cm}
	\label{fig:mnist}
\end{figure*}

\subsection{Handwritten number classification on MNIST}

To evaluate the benefits of the proposed GDL pipeline [cf. Fig.~\ref{fig:classic_and_proposed}(b)], we consider the MNIST classification task \cite{lecun1998gradient}. 
The MNIST dataset consists of $70,\!000$ handwritten digit images from 10 classes corresponding to the different digits. 
The size of each image is $28 \times 28$ pixels. 
We randomly set $55,\!000$ samples as training data, $5,\!000$ as validation data, and $10,\!000$ as testing data.

We consider an $8$ nearest-neighbor representation of each image, where each pixel is connected to its $8$ closest pixels in an unweighted graph of $N \!=\! 784$. 
We consider the raw images $\bby \in \reals^N$ as diffused signals to which we apply a deconvolution step. 
For this step, we use a filter bank with four filters -- uniform, short, medium, and long-range --, each of degree $L=10$ (cf. Section~\ref{Ss:gsp}). Thus, for each input signal $\bby$ we have four (deconvolved) seeding signals $\hat{\bbx}_\text{UR}$, $\hat{\bbx}_\text{SR}$, $\hat{\bbx}_\text{MR}$, and $\hat{\bbx}_\text{LR}$, obtained by solving~\eqref{E:decon_multi_filter}.

For the GNN classifier, we experiment on both a non-parametric GNN and a spline GNN ($Q = 25$), using Graclus as a graph coarsening algorithm for the pooling layer (cf.~Section~\ref{Ss:GDL}). 
The network architecture is a LeNet-5-like CNN \cite{lecun2015lenet} with standard TensorFlow specification given by `GC32-P4-GC64-P4-FC512'. This means that the GNN consists of two graph convolutional layers (GC) with depths of 32 and 64, two pooling layers (P) with a stride of 4, and one fully connected layer (FC) of size 512, configured in the order mentioned above. 

We consider the classification accuracy of both GNNs when applied to four different types of inputs: i) {\bf raw}:~the raw signals (images) $\bby \in \reals^N$, ii) {\bf raw-copy}:~four copies of the raw signals $[\bby, \bby, \bby, \bby] \in \reals^{N \times 4}$, iii) {\bf decon-4}:~the seeding signals $[\hat{\bbx}_\text{UR}, \hat{\bbx}_\text{SR}, \hat{\bbx}_\text{MR}, \hat{\bbx}_\text{LR}] \in \reals^{N \times 4}$, and iv) {\bf decon-1}:~the aggregated seeding signals $\hat{\bbx} = \hat{\bbx}_\text{UR} + \hat{\bbx}_\text{SR} + \hat{\bbx}_\text{MR} + \hat{\bbx}_\text{LR} \in \reals^N$. Notice that the first two approaches fall under the classical setting in Fig.~\ref{fig:classic_and_proposed}(a) with no preprocessing of the raw data while the last two are examples of the proposed deconvolution approach in Fig.~\ref{fig:classic_and_proposed}(b).

The validation accuracy as a function of the training step for both types of GNNs considered is illustrated in Figs.~\ref{fig:mnist}(a)-(b). 
In both cases, the training process for the deconvolved signals is faster and achieves higher values of validation accuracy. 
Notice that even the replicated raw data (raw-copy) achieves lower accuracy levels than the consolidated deconvolved signals (decon-1), indicating that the increase in performance is not attributable to the consideration of larger input signals but rather to the deconvolution step proposed.
Similar conclusions can be obtained when considering the testing accuracy of the trained GNNs; see Fig.~\ref{fig:mnist}(c). Even though the spline GNN achieves better performance than the non-parametric one, for both configurations the deconvolved signals yield a higher accuracy than their raw (non-processed) counterparts.

\section{Conclusions}\label{S:conclusions}
We explored synergies between GDL and GSP that go beyond the area of graph filter design and touch upon the existing work on (blind) graph filter identification.
More precisely, we proposed a pre-processing step for GDL that consists of a deconvolution procedure of the observed signals using a graph filter bank.
This deconvolution step -- formulated as a group-sparse recovery problem -- was shown to improve the empirical classification power of GNNs when compared to the classification of the unprocessed graph signals.
Potential research avenues that follow from the proposed work include the derivation of theoretical guarantees of the deconvolution step to better characterize the classification improvements, and the development of criteria for optimal design of the graph filter bank that drives the deconvolution operation.

\newpage
\bibliographystyle{IEEEtran}
\bibliography{citations}

\end{document}